\documentclass[a4paper,fleqn,usenatbib]{mnras}

\usepackage[T1]{fontenc}
\usepackage{ae,aecompl}

\usepackage{graphicx}	
\usepackage{amsmath}	
\usepackage{amssymb}	

\title[Modelling SXP 5.05]{Modelling the observable behaviour of SXP 5.05}

\author[R. O. Brown et al.]{
R. O. Brown,$^{1}$\thanks{E-mail: rob1g10@soton.ac.uk}
M. J. Coe,$^{1}$
W. C. G. Ho,$^{1, 2, 3}$
A. T. Okazaki$^{4}$
\\
$^{1}$Physics and Astronomy, University of Southampton, Southampton, SO17 1BJ, UK \\
$^{2}$Mathematical Sciences and STAG Research Centre, University of Southampton, Southampton, SO17 1BJ, UK  \\
$^{3}$Department of Physics and Astronomy, Haverford College, 370 Lancaster Avenue, Haverford, PA 19041, USA \\
$^{4}$Faculty of Engineering, Hokkai-Gakuen University, Toyohira-ku, Sapporo, 062-8605, Japan
}

\date{Accepted XXX. Received YYY; in original form ZZZ}

\pubyear{2018}

\begin{document}
\label{firstpage}
\pagerange{\pageref{firstpage}--\pageref{lastpage}}
\maketitle

\begin{abstract} 
SXP 5.05 is a Be/X-ray binary with a neutron star companion located in the Small Magellanic Cloud. It was first detected in 2013, and later that year, SXP 5.05 underwent a massive optical and X-ray outburst.  This outburst dwarfs any other optical event that has been observed for this system during the last 5 years. The large increase in optical brightness of the system implies an increase in the size and density of the Be star's circumstellar disc. The X-ray data show two occultations of the neutron star per orbit and is not consistent with a neutron star passing behind the Be star, and hence the disc is responsible for these occultations. In this paper, we model the outburst of Be/neutron star binary SXP 5.05 as being due to a large increase in mass ejection by the Be star.  The neutron star passes directly through the growing disc, and it is shown that the resulting obscuration can qualitatively explain the observed X-ray behaviour of the system. We find the only way to reproduce the timescales of the observed optical behaviour of the system is to increase the mass ejection substantially for a short time (<2 orbits) and to decrease the viscosity during the event. The general behaviour of the observed X-ray and H$\alpha$ line emission are also reproduced by the model. However, the inferred mass ejection and viscosity needed to produce a sufficiently rapid increase of disc size are both higher than suggested by previous works.
\end{abstract}




\section{Introduction}

Be stars are B spectral type stars that have, or have had at some time, one or more Balmer lines in emission \citep{JascEgre1982}. In comparison to many other main sequence stars, Be stars have high rates of mass outflow, and rotate very quickly \citep{Slettebak1988}. This rapid rotation, in addition to non-radial pulsations, is thought to lead to a diffuse and gaseous circumstellar disc \citep{Riv2000}. This is commonly referred to as a decretion disc. The disc is a dominant mechanism for accretion onto a binary companion, because the stellar wind of a Be star is generally weak. When the companion is a compact object, the system is a High Mass X-ray Binary (HMXB). An upper limit to the decretion rate of particular Be stars was determined to be $\dot{M} = 10^{-8} M_{\odot}$yr$^{-1}$ \citep{Riv2000} but the decretion rate has more recently been refined to the range of $10^{-12} - 10^{-9} M_{\odot}$yr$^{-1}$ \citep{Vieira2015}. The mass-loss rate of the Be star for observable (and likely more dense) discs has been shown to be $\dot{M} = 10^{-10} M_{\odot}$yr$^{-1}$ by \citet{Rimulo2018}. \citet{Ghor2018} find that the lightcurve of $\omega$ CMa can be described by assuming mass ejections as high as $\sim 4 \times 10^{-7}M_{\odot}$yr$^{-1}$. It should be noted that the mass ejection described by \citet{Ghor2018} is higher than the mass-loss/decretion rate due to the material that is immediately reaccreted onto the Be star \citep{Haubois2012}. In this paper, mass ejection is discussed and this is larger than the amount of material lost to the disc due to the accretion of the Be star.

Be/X-ray binaries are the largest population of observable HMXBs \citep{RapHeu1982, HeuRap1987, Coleiro2013}. The varying size of the disc, coupled with the interaction of a compact object leads to a variety of observable effects. Some of these, such as superorbital modulation and giant outbursts, are not well understood. Comprehending these phenomena can lead to a better understanding of the extreme physics of neutron stars and black holes (see \citet{Reig2011} for review).

SXP 5.05 was first detected via INTErnational Gamma-Ray Astrophysics Laboratory (INTEGRAL) observatory \citep{Coe2013} as a bright X-ray source based in the Small Magellanic Cloud. Due to the eclipsing nature of the binary, its orbital period was accurately measured as 17.13 days.

\citet{Coe2015} considered the X-ray behaviour of SXP 5.05 in further detail, allowing for the determination of an accurate orbital model solution. The orbital eccentricity is found to be 0.155 and the orbital period is confirmed to be 17.13 days. It is calculated by fitting to a simple spin-up model and a radial velocity model simultaneously using Doppler-shifted X-ray pulsations. The initial attempts resulted in poor fits. A highly variable accretion rate caused by a clumpy wind was suggested to be the cause. The neutron star is occulted not by the Be star but by some extended structure which is interpreted as the circumstellar disc of the Be star. A simple 2-dimensional model is proposed that involves the neutron star ploughing through the circumstellar disc that is perpendicular to the orbital plane.

\begin{figure}
	\centering
	\includegraphics[width=.5\textwidth]{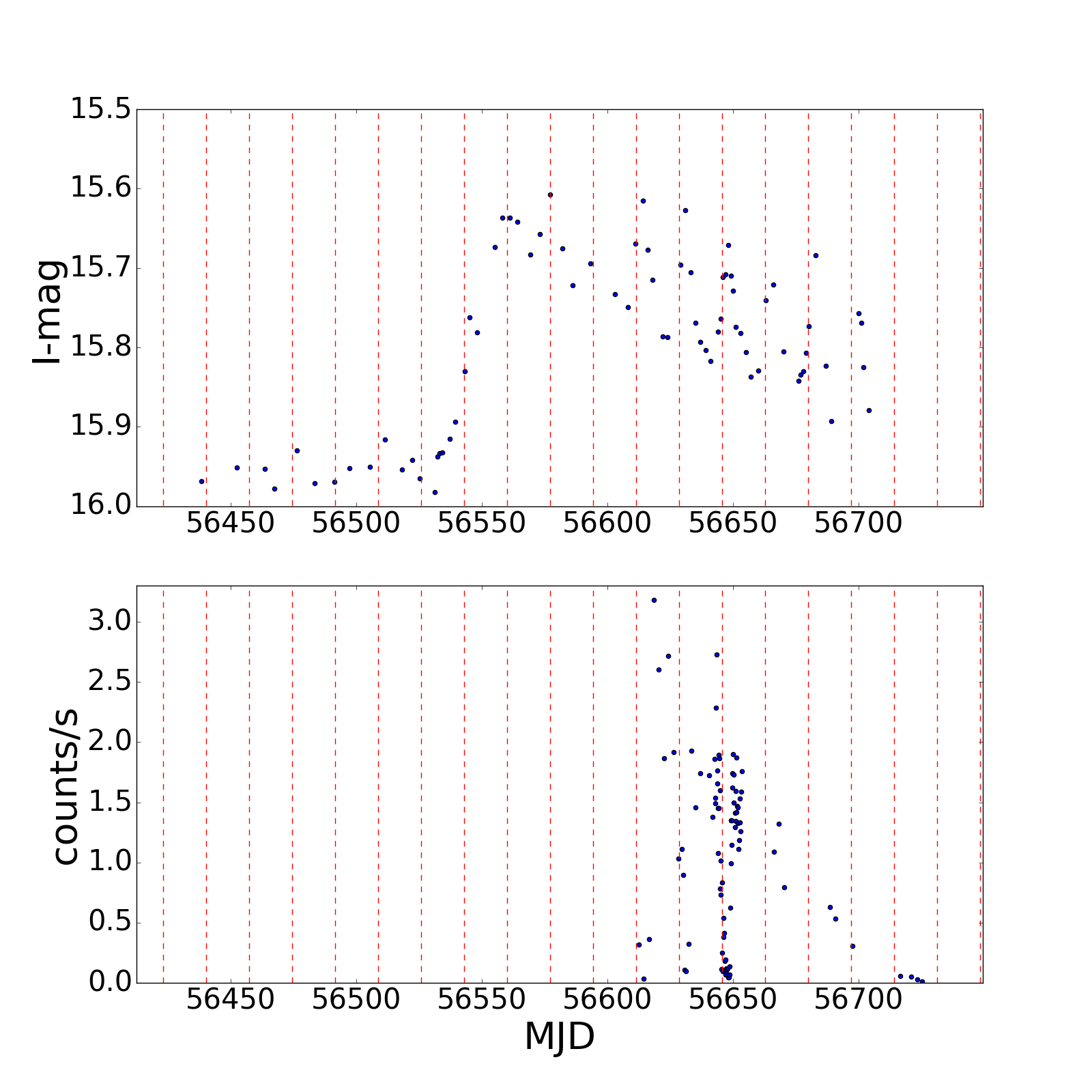}
	\caption{Observational data of SXP 5.05 over a period of 200 days, taken from \citet{Coe2015}. Red dashed lines indicate the neutron star's periastron. Top: OGLE I-band measurements. Bottom: Swift X-ray flux measurements.}
	\label{fig:RealData}
\end{figure}

Figure \ref{fig:RealData} shows simultaneous optical and X-ray observations of SXP 5.05 over a 200 day period that are shown in \citet{Coe2015}. Note that in the optical band, there is often a minimum before periastron and often a maximum following it. There is a large increase in the size of the disc that takes place over $\sim$3-4 orbits. The X-ray data show an extreme minimum following every periastron for approximately one fifth of an orbital period (see Section \ref{sec:XrayLuminosity}). The maximum value of X-ray flux occurs shortly before periastron.

In this paper, the effects of a neutron star travelling directly through the disc are investigated. Simulation results are used to explain the behaviour of the observational data shown in Figure \ref{fig:RealData}. Section \ref{sec:Simulations} states the properties of all the simulations discussed within the paper. In Section \ref{sec:Optical}, the visible area of the disc is compared to the observed optical data, with the aim of matching the timescales over which the I-band magnitude rises and falls during the outburst. This part of the paper is used to restrict the parameters and determine the best fitting systems. Section \ref{sec:Obscuration} discusses the column density of neutral hydrogen that obscures the neutron star for the restricted sample of simulations. The fitting of both the optical timescales in the previous section and the obscuration allows for one simulation to be chosen as the best match to the observational data. Section \ref{sec:XrayLuminosity} uses the predicted obscuration to determine the X-ray behaviour of the neutron star from an observer's point of view. Section \ref{sec:Ha} shows the evolution of the H$\alpha$ line emission shape for the Be star's circumstellar disc. Section \ref{sec:Conclusions} discusses the comparison between the simulation and observational data and it is considered whether the passage of the neutron star through the disc is an explanation for the X-ray occultations in SXP 5.05.

\section{Simulations} \label{sec:Simulations}

A 3D smoothed particle hydrodynamics code is used to simulate Be/X-ray binaries. The Be star's decretion disc is modelled by an ensemble of particles each of mass $\sim 10^{-15}$M$_{\odot}$. The disc is assumed to be isothermal with a temperature of $T = 0.6 T_{eff}$ for simplicity \citep{PoeMarl1982, vanKer1995, CarcBjor2006}. The particles are injected into the disc with Keplerian velocity at a random azimuthal angle at 1.05 stellar radii from the centre of the Be star. They are placed at a random small distance from the equatorial plane. The Be star and compact object are modelled using sink particles. For further details on the code, see \citet{Brown2018} and the references therein.

\begin{figure}
	\centering
	\includegraphics[width=.4\textwidth]{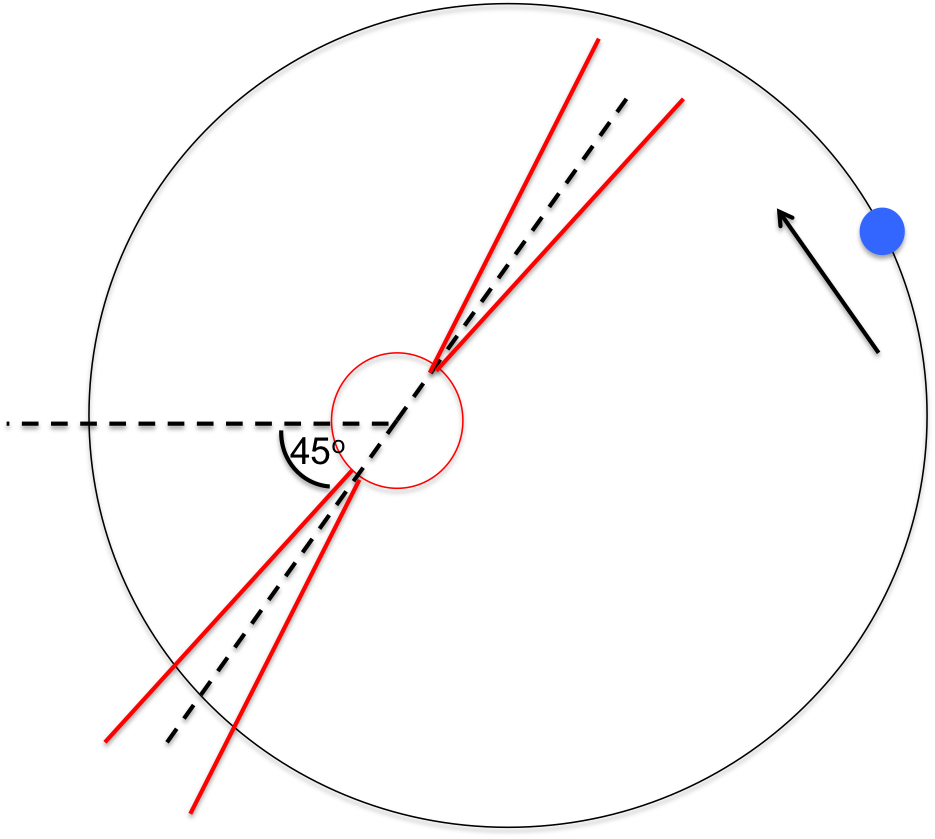}
	\includegraphics[width=.4\textwidth]{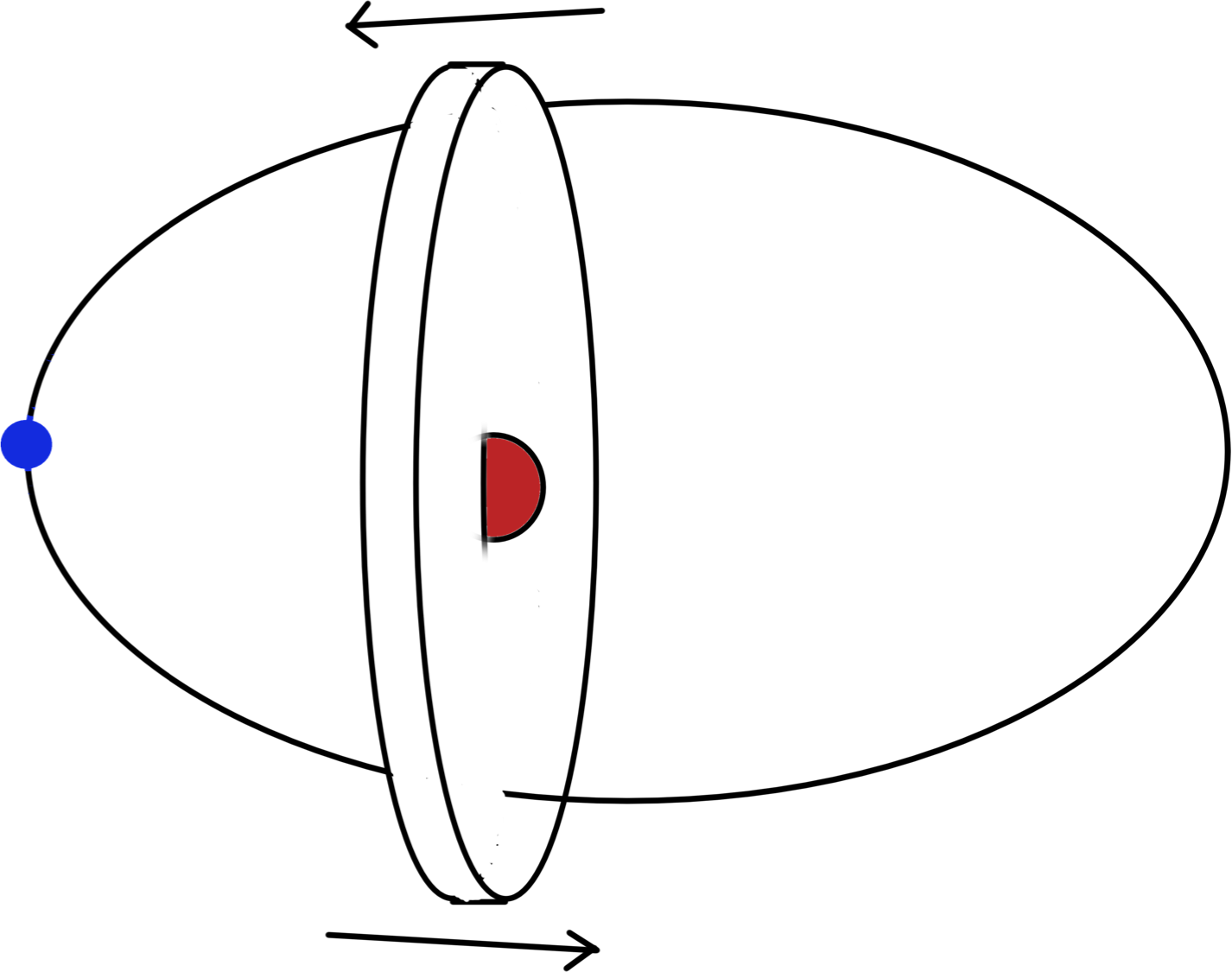}
	\caption{Top: An illustration of the proposed geometry of the Be/X-ray binary SXP 5.05. The red circle is the Be star and the connected red lines indicate the flaring circumstellar disc. The blue circle is the orbiting neutron star. Dashed lines depict the 45$^{\circ}$ angle of the disc to periastron. Bottom: An illustration of the suggested observer's view of SXP 5.05. The red circle surrounded by a disc and the blue circle represent the Be star and neutron star respectively. The neutron star orbits anti-clockwise, as shown by the arrows, and is at periastron.}
	\label{fig:SystemGeometry}
\end{figure}

The simulations have an initial mass ejection rate of $10^{-11}$ M$_{\odot}$yr$^{-1}$ and are evolved until they reach equilibrium, i.e. when the number of particles in the disc around an orbit changes by less than 1$\%$ for more than 5 orbits. The simulations shown in this paper contain 20,000 to 50,000 particles at equilibrium. Following this, the mass ejection rate is increased and the system evolved further. Simulations are performed with increased mass ejection rates of $10^{-10}, 10^{-9}, 10^{-8}, 10^{-7}$, $10^{-6}$ and $10^{-5}$ M$_{\odot}$yr$^{-1}$. Sections \ref{sec:viscosity} and \ref{sec:ejection} describe simulations where the mass ejection of the Be star is increased and then kept at the increased value. The simulation that is discussed for the remainder of the paper has the Be star's mass ejection rate increased from $10^{-11}$ M$_{\odot}$yr$^{-1}$ to $10^{-5}$ M$_{\odot}$yr$^{-1}$ instantaneously and then the mass ejection rate is decreased in steps over time to the original value of $10^{-11}$ M$_{\odot}$yr$^{-1}$ (see Section \ref{sec:visceject}). When the mass ejection of the Be star is increased, the disc grows in size temporarily and then shrinks until it reaches a state of equilibrium (see Section \ref{sec:ejection} for further detail). During this process, the disc grows larger than the orbit of the neutron star (see Figure \ref{fig:SystemGeometry}). This causes the neutron star to pass directly through the disc at $\sim10$ stellar radii away from the centre of the Be star.

The Shakura-Sunyaev viscosity parameter of the simulation is also varied to best fit the data. The assumed values lie between $\alpha = 0.1$ and $1.5$. Note that $\alpha=0.63$ and $\alpha=0.26$ have been theoretically determined as the median values of the viscosity parameter during build up and dissipation respectively \citep{Rimulo2018} and \citet{Ghor2018} use viscosities varying between $\alpha=0.1$ and $\alpha=1.0$ to successfully model the lightcurve of $\omega$ CMa.

All systems have a Be star of mass 13$M_{\odot}$ and radius 7$R_{\odot}$, as determined in \citet{Coe2015}, and a neutron star of mass 1.4$M_{\odot}$ and radius 10km. The Be star's circumstellar disc is at a 45$^{\circ}$ angle to periastron (illustrated in Figure \ref{fig:SystemGeometry}) and is perpendicular to the orbital plane, thus making the rotation of the disc arbitrary when considering accretion onto the neutron star. 

\subsection{Changes in mass ejection}

When the mass ejection rate of a Be star is modified, the material in the disc shifts to assume a new density profile that remains in equilibrium with the new mass ejection rate. One such example of this has been shown by \citet{Okazaki2002} where a Be star with no circumstellar disc is given a constant mass ejection rate. The disc is shown to grow into a state of equilibrium. The time it takes for the density to reach equilibrium increases with radius. \citet{Haubois2012} showed that two different regions are created when the mass ejection rate of the Be star (with a built up circumstellar disc) is reduced to zero. The first of these is an inner region of material flowing towards the Be star due to the material being accreted back onto the Be star. The other outer region continues to flow outwards and behaves as if the ejection rate of the Be star is unchanged. The radius that separates these two regions is known as the stagnation point and moves outwards with time until the entire disc is flowing inwards.

In this paper, the mass ejection rate of the Be star is increased instantaneously by at least an order of magnitude. This creates a region of higher density at the equator of the Be star that propagates outwards. Much like the stagnation point, there is a radial separation between the two regions that moves away from the central star. However, unlike the case where the ejection rate is reduced, both regions are moving outwards. This leads to an increase in the size of the disc that is proportional to the size of the change in mass ejection. The inner and outer regions are separated by a large gradient in density.

In the case of a Be star with a binary companion, the disc can grow larger than the radius at which it would usually be limited to, i.e. the truncation radius \citep{Okazaki2002}. The disc can initially grow rapidly due to the phenomena described above but when the overall density profile of the disc comes closer to equilibrium, the disc shrinks quickly again due to the truncation of the binary partner. This means that the Be star's circumstellar disc can initially grow and then shrink to its original size even when the mass ejection rate remains constant (see Figure \ref{fig:VisibleDiscArea_effects}).

\section{Optical behaviour} \label{sec:Optical}

Variations in the optical brightness of SXP 5.05 are related to the properties of the circumstellar disc. Hence, the visible area of the disc is calculated and the behaviour of this quantity is used as a preliminary fit to the observational data. The aim is to reproduce the rate of increase and decrease of I-band flux, as shown in Figure \ref{fig:RealData}. Variation over time of both mass ejection and artificial viscosity is investigated. Note that there are only observational X-ray data for the final part of the OGLE data. 

The visible area of the disc is calculated straightforwardly by placing a grid on the system oriented to the observer's line of sight. The addition of the number of grid squares that contain disc matter provides the resultant area. As this is a simplified method and the emission of the disc is not being modelled accurately, the maximum and minimum of the visible disc area is scaled to the observed I-band flux. Thus, absolute values of flux are not comparable, only the times of the changes. 

This treatment assumes that the entirety of the disc is optically thick. \citet{Vieira2015} show that the Be star's disc is optically thick up to a given radius that is wavelength dependent. This region is known as the pseudo-photosphere and its radius, $\overline{R}$, is a function of the stellar parameters as follows

\begin{equation}
	\overline{R} \propto \rho_{0}^{2/(2n-\beta)} \lambda^{(2+u)/(2n-\beta)},
	\label{eq:VieiraRbar}
\end{equation}

\noindent where $\rho_{0}$ is the base gas density of the Be star's circumstellar disc, $\lambda$ is the given wavelength, $\beta$ is the disc flaring exponent and $u$ is given by

\begin{equation}
	u = \frac{d \hspace{1mm} ln(   g_{\mathrm{ff}} + g_{\mathrm{bf}}   )}      { d \hspace{1mm} ln \lambda}.
	\label{eq:VieiraU}
\end{equation}

\noindent $g_{\mathrm{ff}}$ and $g_{\mathrm{bf}}$ are free-free and bound-free gaunt factors, respectively. \citet{Vieira2015} show that a disc with a base gas density of $\rho_{0} = 8.4 \times 10^{-11}$ g$\hspace{0.4mm}$cm$^{-3}$ has a pseudo-photosphere that extends to $\sim 3.5$ stellar radii for a wavelength of $\lambda=2 \mu$m (the I-band includes $\sim 0.1 - 1 \mu$m). The base gas density of the simulations during the event are orders of magnitude higher than this (up to $\sim 10^{-7}$ g$\hspace{0.4mm}$cm$^{-3}$) and hence a scaled $\overline{R}$ is certainly larger than the maximum radius that the disc reaches ($\sim 20$ stellar radii). However, there are two issues with this assumption. Firstly, the starting point of the simulation has a base gas density comparable to $8.4 \times 10^{-11}$ g$\hspace{0.4mm}$cm$^{-3}$ and hence $\overline{R}$ does not extend to the edge of the disc in this case. The other issue is that during the event, the density profile of the disc is not well behaved and so it is difficult to consider exactly where the pseudo-photosphere ends. The assumption that the entirety of the disc is equally emissive in the I-band is a simplification and using radiative transfer methods would yield more accurate results.

\subsection{Effect of viscosity} \label{sec:viscosity}

The top plot in Figure \ref{fig:VisibleDiscArea_effects} shows the visible disc area for four simulations of varying $\alpha$. The values shown are relative to the visible disc area at moment the Be star's mass ejection is increased and hence do not represent the size of each individual disc at equilibrium. The area of the disc with a viscosity parameter of $\alpha = 0.1$ is almost two times larger than the disc with $\alpha = 1$.

The rate of disc growth decreases with larger viscosity parameters. However, discs with larger viscosity are found to decrease in size faster after reaching a maximum, implying the need for a lower viscosity when the disc is shrinking. The rate of disc recession remains very close to the rate of growth for all values of $\alpha$. Reducing the viscosity alone is not sufficient to reproduce the slower decline of the second half of the data. 

There is a delay between the increase in ejection and the growth of the visible area of the disc. This is due to the viscosity timescale of the disc, i.e. the time it takes for the region of higher density to propagate through the disc \citep{King2013}. The viscosity timescale of the disc is given by

\begin{equation}
	\tau_{\mathrm{visc}} =  \frac{r^{2}}{\nu},
	\label{eq:viscosityTimescale}
\end{equation}

\noindent where $r$ is the distance the higher density region has travelled. The viscosity of the fluid, $\nu$, is given by

\begin{equation}
	\nu =  \frac{\alpha c_{s}^{2}}{\Omega_{\mathrm{K}}(r)},
	\label{eq:viscosity}
\end{equation}

\noindent where $\alpha$ is the Shakura-Sunyaev viscosity parameter, $c_{s}$ is the speed of sound in the fluid and $\Omega_{K}$ is the Keplerian velocity at the radius $r$. The decretion disc in the simulations extends to 5-10$R_{*}$ and $\alpha$ varies from 0.1 to 1. The speed of sound calculated from the simulations is $\sim 10$km s$^{-1}$. Thus, the time taken for a feature to propagate from the edge of the Be star to the edge of the disc is of the order of hundreds of days for systems with $\alpha \sim 1$. This timescale is closer to a thousand days for the lower values of $\alpha$. However, the growth of the disc is faster than the viscosity timescale because the newly ejected material displaces the outer disc. Therefore, by the time the higher density region reaches the edge of the initial disc, it has already grown considerably.

\subsection{Effect of mass ejection} \label{sec:ejection}

\begin{figure}
	\centering
	\includegraphics[width=.5\textwidth]{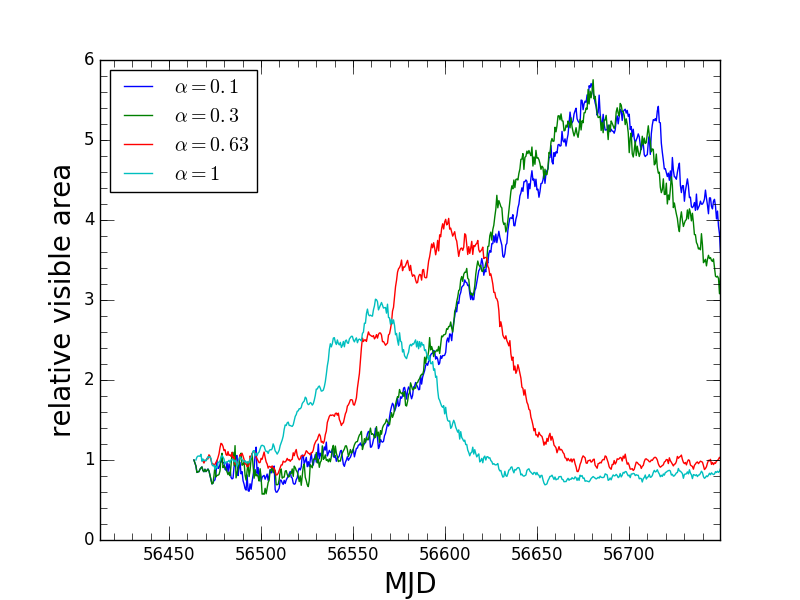}
	\includegraphics[width=.5\textwidth]{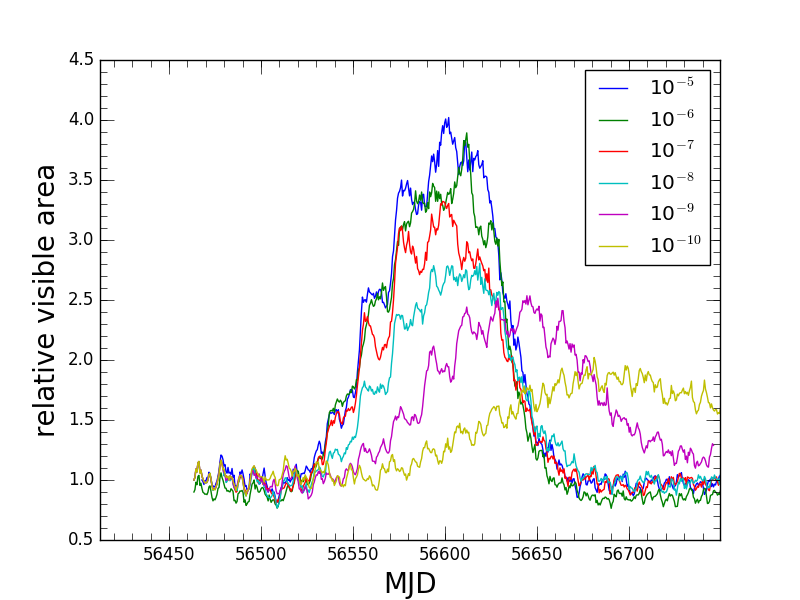} 
	\caption{Top: The visible area of the disc against time for four systems with viscosities parameters of $\alpha = 0.1, 0.3, 0.63$ and $1$. It is relative to the visible area at the time when the mass ejection is increased (the initial point of each line). All four systems have had the mass ejection of the Be star increased from $10^{-11}$ to $10^{-5}$  M$_{\odot}$yr$^{-1}$. Bottom: The visible area of the disc against time for systems with increased Be star mass ejection of $10^{-10}, 10^{-9}, 10^{-8}, 10^{-7}$, $10^{-6}$ and $10^{-5}$ M$_{\odot}$yr$^{-1}$. It is relative to the visible area at the time when the mass ejection is increased (the initial point of each line). All systems have a viscosity parameter of $\alpha = 0.63$. }
	\label{fig:VisibleDiscArea_effects}
\end{figure}

The bottom plot in Figure \ref{fig:VisibleDiscArea_effects} shows the visible disc area with time for six simulations that have had their mass ejections increased from $10^{-11}$ M$_{\odot}$yr$^{-1}$. Greater increases in mass ejection lead to a faster rise and fall in the visible area of the disc, similar to viscosity. Smaller increases of mass ejection require a much longer time for the visible disc area to fall than the time for it to rise. None of the systems shown in the bottom of Figure \ref{fig:VisibleDiscArea_effects} exhibit both the rapid increase and slower decline that were observed in SXP 5.05. Hence, a more complex variation of mass ejection over time is necessary.

The behaviour of the disc is extremely sensitive to both the duration and magnitude of any mass ejection variation. Maintaining an increased mass ejection for more than two orbits leads to the approximately symmetric behaviour seen in the top of Figure \ref{fig:VisibleDiscArea_effects}. We find the only way to replicate the times of both the rise and fall of the optical data is to have a short and large initial increase in mass ejection, after which, the mass ejection falls again.

\subsection{Combining viscosity and mass ejection} \label{sec:visceject}

One would reasonably expect the mass ejection to decrease as a function of time after a large outburst. Thus, the mass ejection of the Be star in this section is decreased in steps until it reaches the original rate of $10^{-11}$ M$_{\odot}$yr$^{-1}$. The time between steps is assumed to be approximately a binary period and the steps are each an order of magnitude in M$_{\odot}$yr$^{-1}$. This is combined with a single change in the viscosity parameter from $\alpha=1.5$ to $\alpha=0.1$ during the event. In order to recover the rapid rise to maximum visible disc area, the viscosity is be changed a approximately five orbits after the mass ejection is increased.

\begin{figure}
	\centering
	\includegraphics[width=.5\textwidth]{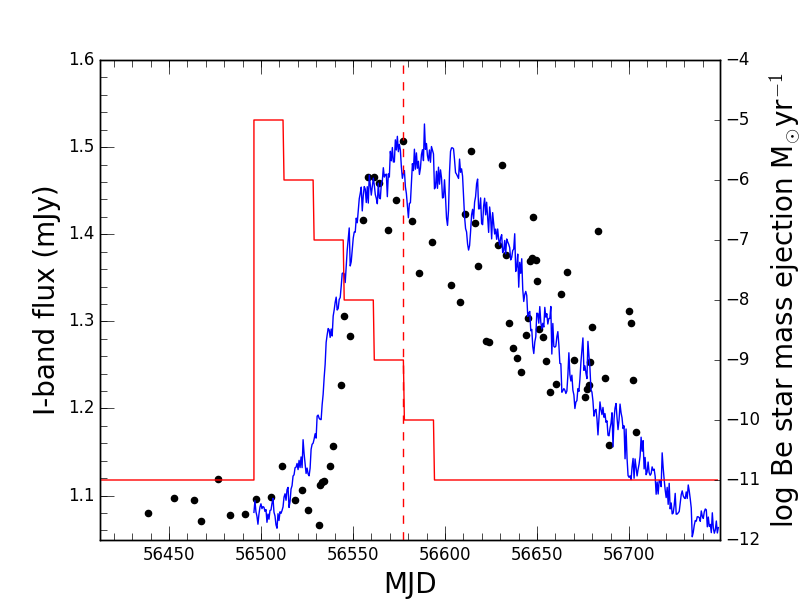}
	\caption{OGLE flux (black points) and estimated I-band flux (blue line) for best-fitting simulation as function of time. The mass ejection of the system is initially increased from $10^{-11}$ M$_{\odot}$yr$^{-1}$ to $10^{-5}$ M$_{\odot}$yr$^{-1}$ and then decreased by an order of magnitude every $\sim P_{orb}$ until the ejection rate has returned to $10^{-11}$ M$_{\odot}$yr$^{-1}$. This is illustrated by the solid red line. The vertical dashed red line indicates when the viscosity parameter is changed from $\alpha=1.5$ to $\alpha=0.1$. The OGLE data is shown by the black points.}
	\label{fig:opticalBestFit}
\end{figure}

Figure \ref{fig:opticalBestFit} shows the visible disc area compared against the observational optical data for the best fitting simulation. The system begins at equilibrium, with a Be star mass ejection of $10^{-11}$ M$_{\odot}$yr$^{-1}$. The Be star's mass ejection is then increased to $10^{-5}$ M$_{\odot}$yr$^{-1}$ and the viscosity parameter is set to $\alpha=1.5$. After approximately one orbit the mass ejection is then decreased by an order of magnitude every orbit until it reaches the original rate of $10^{-11}$ M$_{\odot}$yr$^{-1}$. Approximately five orbits after the initial increase in mass ejection, viscosity is changed to $\alpha=0.1$ (the same time that ejection is changed to $10^{-10}$ M$_{\odot}$yr$^{-1}$). Although this is the simulation that best fits the overall evolution of the I-band flux, it does not fit the orbital modulation. Firstly, the model does not consistently brighten shortly after every periastron like the observational data does. The simulated optical flux also does not reproduce the large amplitude of the variations in the second half of the data. This implies the method used for modelling the optical lightcurve struggles to model the short-term variations.

\begin{figure}
	\centering
	\includegraphics[width=.44\textwidth]{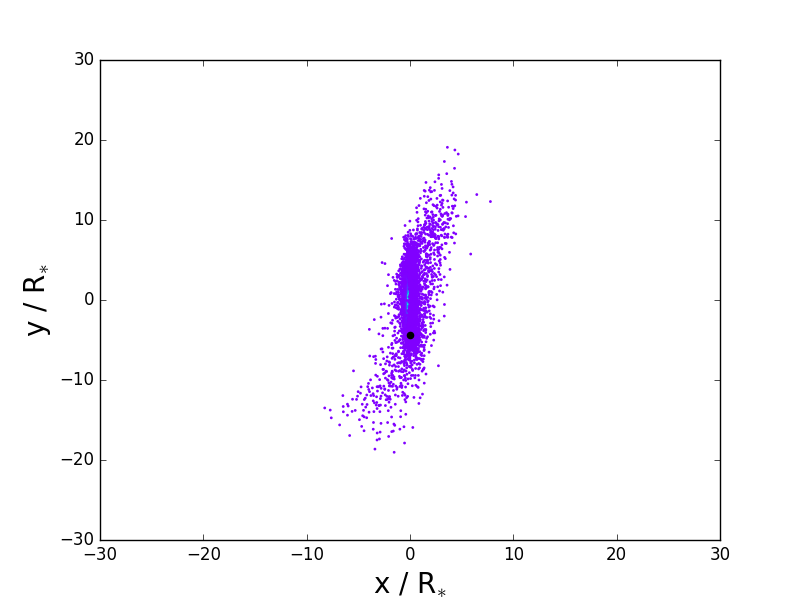}
	\includegraphics[width=.44\textwidth]{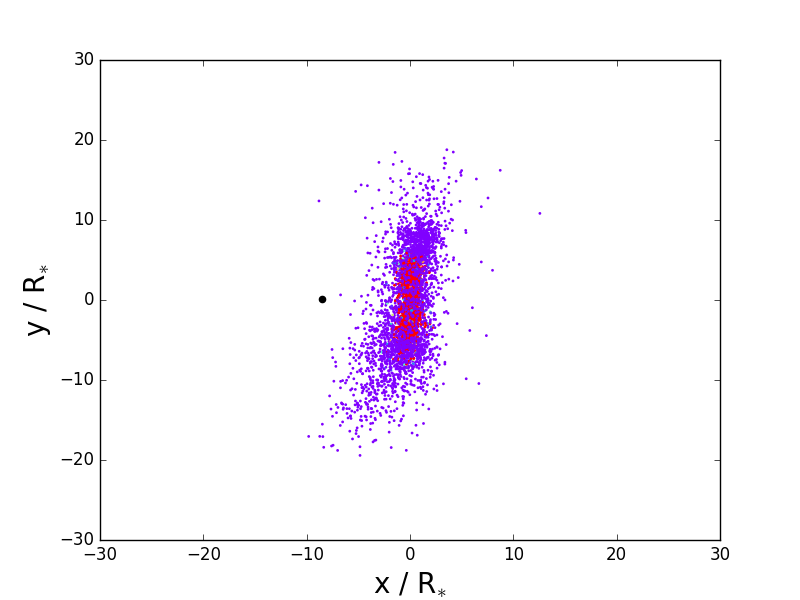}
	\includegraphics[width=.44\textwidth]{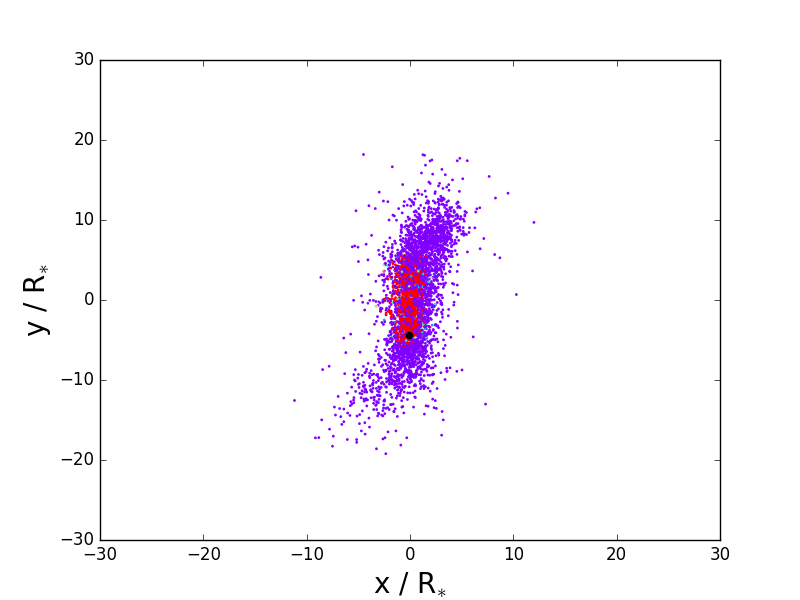}
	\caption{Snapshots of the disc and neutron star for the best fitting system shown in Figure \ref{fig:opticalBestFit}. The top plot shows the system at the moment when the mass ejection is increased and the middle and bottom plots show the system $\sim 3.8$ and $\sim 4$ orbits after the first snapshot, respectively. The three snapshots from top to bottom show the system at binary phases of $\sim 0.2$, $\sim 0$ and $\sim 0.2$, repsectively. Red points shows the inner, higher density region of the disc created by the increased mass ejection and purple points show the lower density region that remains from the initial disc. The solid black circle shows the position of the neutron star that lies in front of the disc. This plot is shown in the assumed observer's line of sight.}
	\label{fig:areaBestFitSystem}
\end{figure}

Figure \ref{fig:areaBestFitSystem} is an illustration of the best fitting system, shown in Figure \ref{fig:opticalBestFit}, at two times. The first plot shows the moment the Be star's mass ejection is increased and the second is four orbits after. As the disc grows in size, the outer regions interact more closely with the neutron star. Due to the disc's misalignment to the orbital plane, the outer regions bend out of the plane of the disc (see Figure \ref{fig:areaBestFitSystem}). This gives rise to a greater rate of increase in visible disc area over the entire event. The larger disc also yields a thicker edge and, as the disc is $\sim 30^{\circ}$ away from edge on, is another contributing factor to the increasing visible area. 

The bending of the disc could also explain the brightening just after periastron seen in the optical data (see Figure \ref{fig:RealData}). The second and third snapshot in Figure \ref{fig:areaBestFitSystem} show the best fitting simulation at binary phases of $\sim 0$ and $\sim 0.2$. The disc after periastron curves at the top and bulges yielding a larger area. However, because these changes in the size of the disc are small ($\sim 6\%$) it is hard to discern the changes by eye.

The expected range of mass ejection for Be stars is $10^{-12} - 10^{-9} M_{\odot}$yr$^{-1}$ \citep{Vieira2015}, with typical values for the mass ejection of observable stars being $10^{-10}$ M$_{\odot}$yr$^{-1}$ \citep{Rimulo2018}. The outburst that SXP 5.05 undergoes is dependent on both viscosity and mass ejection, and thus, there is more than one possible combination of these quantities that can explain it. However, it is required that either viscosity, mass ejection or both are higher than theory suggests for the observed optical timescales to occur. The system shown in Figure \ref{fig:opticalBestFit} is used for the data in the remainder of this paper.

\section{Disc eccentricity} \label{sec:Eccentricity}

The Kozai-Lidov mechanism is important to consider because of the large inclination of the disc to the orbital plane \citep{Martin2014}. This means that the disc could be largely eccentric and even undergo fragmentation in extreme circumstances \citep{Fu2017}. Using Equation 3 from \citet{Martin2014}, the period of the Kozai-Lidov mechanism for this system is of the order of $\sim 100$ P$_{\mathrm{orb}}$ and the requirement for equilibrium is <10 P$_{\mathrm{orb}}$. 

\begin{figure}
	\centering
	\includegraphics[width=.5\textwidth]{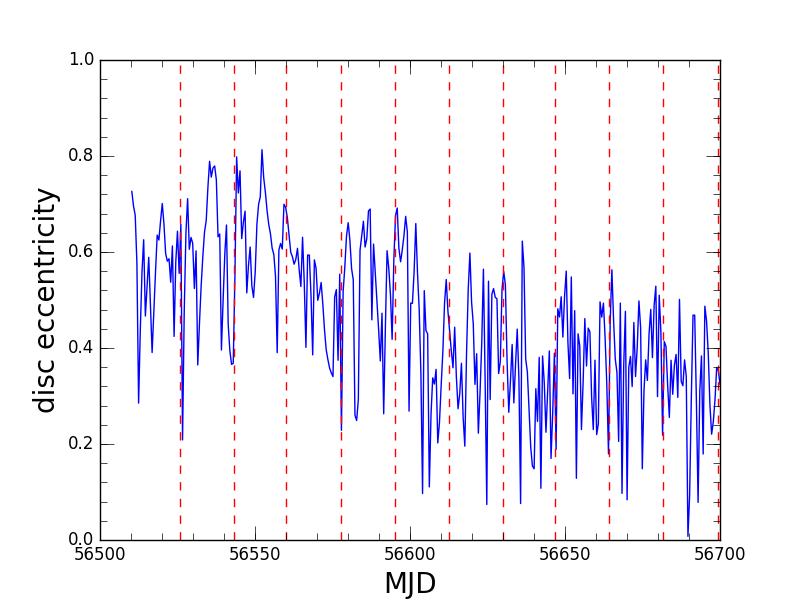}
	\caption{The eccentricity of the Be star's circumstellar disc with time for the best fitting system shown in Figure \ref{fig:opticalBestFit}. The red dashed lines denote periastron.}
	\label{fig:discEcc}
\end{figure}

Figure \ref{fig:discEcc} shows the eccentricity of the Be star's circumstaller disc with time. Before the mass ejection of the Be star is increased, the eccentricity of the disc is very high, reaching a maximum of $\sim 0.8$ and then it decreases over the event. Due to the extreme inclination of the disc, the maximum eccentricity possible is $e \sim 1$. The eccentricity is close to $0.8$ when the mass ejection is increased and thus, the Kozai-Lidov mechanism has started to take effect. When the mass ejection is increased, a large number of new particles flow outward from the star with circular orbits around the Be star. This suppresses the eccentricity of the disc as seen in Figure \ref{fig:discEcc}.

\begin{figure}
	\centering
	\includegraphics[width=.22\textwidth]{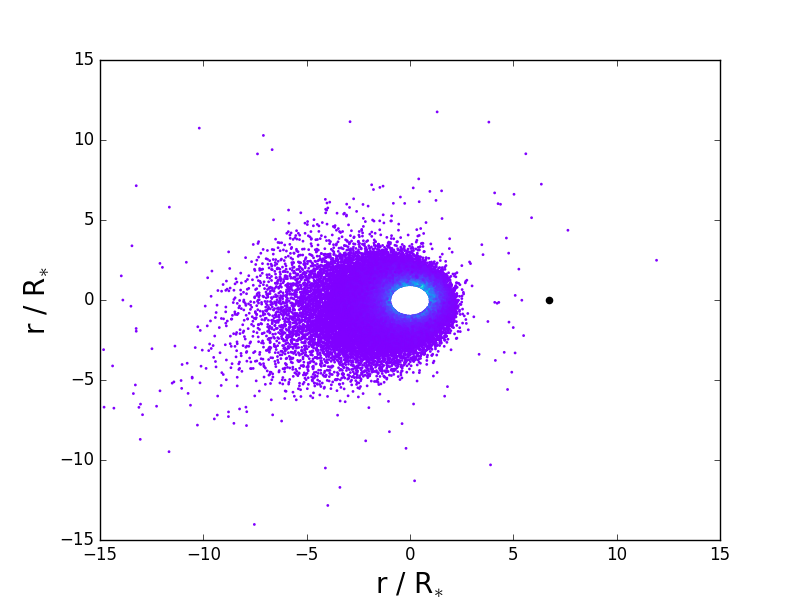}
	\includegraphics[width=.22\textwidth]{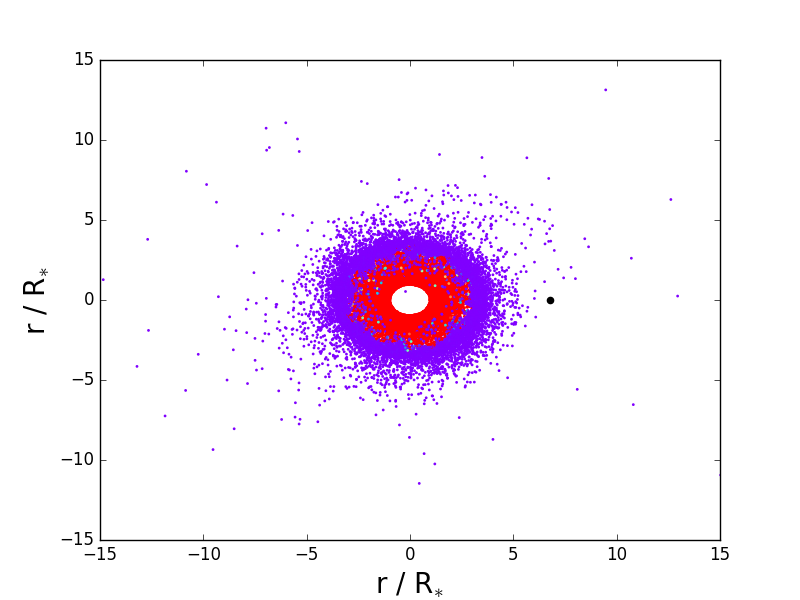}
	\caption{Snapshots of the disc and neutron star in the Be/X-ray binary with a viscosity of $\alpha=0.1$, which has had its mass ejection changed from $10^{-11}$ M$_{\odot}$yr$^{-1}$ to $10^{-5}$ M$_{\odot}$yr$^{-1}$ (a system contained in Figure \ref{fig:VisibleDiscArea_effects}). The left plot shows the system at the moment when the mass ejection is increased and the plot on the right shows the system four orbits after that. Red points shows the inner, higher density region of the disc created by the increased mass ejection and purple points show the lower density region that remains from the initial disc. The solid black circle shows the position of the neutron star and the Be star is reprensented by the white space at the centre of the disc. The plots are in the plane of the disc.}
	\label{fig:discEccPic}
\end{figure}

The variation in eccentricity is demonstrated visually in Figure \ref{fig:discEccPic}. The change in eccentricity is greater for discs with a smaller viscosity parameter and hence the system shown has a viscosity parameter of $\alpha=0.1$. The shape of the disc at the starting point (the moment the mass ejection is changed) is extremely eccentric and within four orbits has circularised considerably. This effect is much less visible in the best fitting system shown in Figure \ref{fig:opticalBestFit}. This rapid circularisation is dependent on the viscous forces in the disc, as has been seen previously \citep{Carciofi2011}. The Kozai-Lidov mechanism has been shown to behave normally in hydrodynamical discs \citet{Martin2014} such as the model used in this paper. However, under such extreme events like the outburst discussed here, the mechanism is disrupted. 

\section{Neutron star obscuration} \label{sec:Obscuration}

The X-ray behaviour of SXP 5.05 can be explained by a disc that grows larger than the neutron star's orbit. If the neutron star passes directly through the disc, any X-ray emission that is visible to an observer will be obscured by the material along the line of sight. The column density of neutral hydrogen, $N_{\mathrm{H}}$, is shown in Figure 19 of \citet{Coe2015}. To mimic these observations, the simulation data should show a peak in obscuration at a binary phase of $\sim 0.1$ with values in the range $N_{\mathrm{H}} \sim 10^{21} - 10^{24}$cm$^{-2}$. 

The column density of neutral hydrogen atoms is calculated using the Saha ionisation equation in the following form

\begin{equation}
	\frac{x^{2}}{1-x} = \frac{1}{n} \left( \frac{2\pi m_{e}kT}{h^{2}} \right)^{3 / 2} e^{-13.6eV / kT} ,
	\label{eq:ionFrac}
\end{equation}

\noindent where $x$ is the fraction of ionised hydrogen, $n$ is the column density of gas particles and $T$ is the temperature of the gas. The column density is calculated by integrating in a cylinder along line of sight to the neutron star. As the neutron star is only obscured by elements of the disc at $\sim 10$ stellar radii or greater, a temperature of $T = 0.6 T_{eff}$ is adopted. The disc is assumed to be composed entirely of hydrogen and a column density of $N_{\mathrm{H}} \sim 10^{21.5}$ is used for the interstellar medium, to match the data in \citet{Coe2015}.

\begin{figure}
	\centering
	\includegraphics[width=.5\textwidth]{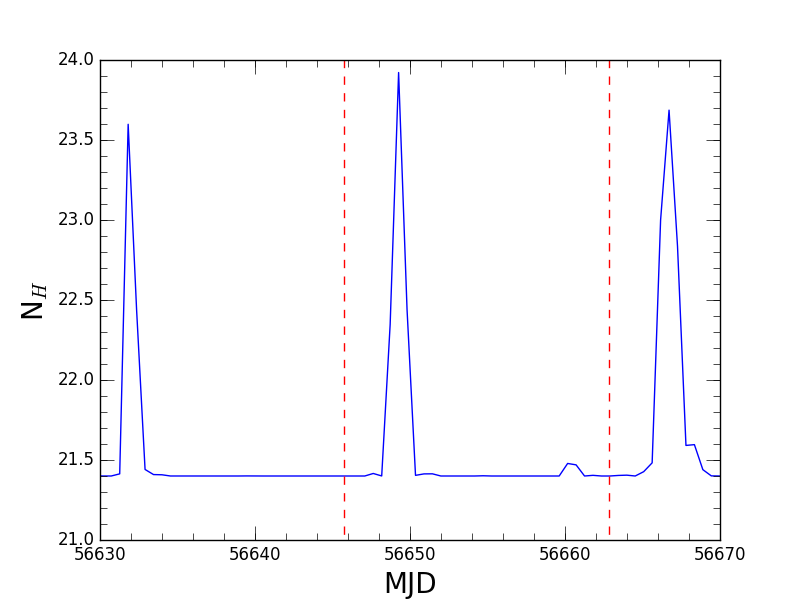}
	\caption{The column density of neutral hydrogen obscuring the neutron star against time for the best fitting system shown in Figure \ref{fig:opticalBestFit}. The red dashed lines denote periastron.}
	\label{fig:NSObscuration}
\end{figure}

Figure \ref{fig:NSObscuration} shows the amount of disc material that obscures the neutron star along a viewing angle illustrated in Figure \ref{fig:SystemGeometry}. There are two occultations per orbit, one just after periastron at a binary phase of $\sim 0.1$ and one preceding periastron which is negligible in comparison. The duration of the larger obscuration after periastron is approximately one fifth of an orbit. Therefore, the time and duration of the peak obscuration in Figure \ref{fig:NSObscuration} agrees with the observed values shown in \citet{Coe2015}. $N_{\mathrm{H}}$ in the simulation remains inside the observational data's range of $\sim 10^{21} - 10^{24}$cm$^{-2}$.

\section{Neutron star X-ray luminosity} \label{sec:XrayLuminosity}

The X-ray luminosity of the neutron star is calculated from its captured mass as follows

\begin{equation} 
L_{X} = \frac{G M_{X} \dot{M}} {R_{X}},
\label{eq:LX}
\end{equation}

\noindent where $\dot{M}$ is the rate of mass capture and $M_{X}$ and $R_{X}$ are the compact object's mass and radius respectively. The mass capture rate is calculated directly from the model by monitoring the number of simulation particles that fall inside the neutron star's radius (each particle has a mass of $\sim 10^{-15}$M$_{\odot}$). The luminosity predicted by the simulations is converted into an observed flux by assuming a distance of 60 kpc \citep{Scowcroft2016}. Using the amount of obscuring matter (as calculated in Section \ref{sec:Obscuration}) the expected counts per second can be computed using WebPIMMS (found at \url{https://heasarc.gsfc.nasa.gov/cgi-bin/Tools/w3pimms/w3pimms.pl}). The energy range for WebPIMMS is set to 1-10 keV and the Galactic column density of hydrogen is assumed to be $\sim 10^{21.5}$cm$^{-2}$. A photon index of 1.53 is used, as determined in \citet{Coe2015}. 

\begin{figure}
	\centering
	\includegraphics[width=.5\textwidth]{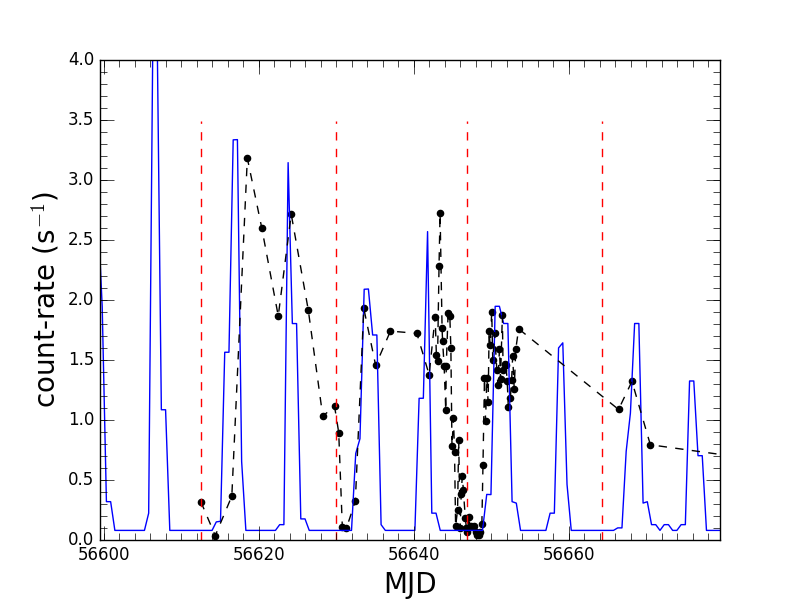}
	\caption{The predicted X-ray counts per second of the neutron star calculated for the best fitting system shown in Figure \ref{fig:opticalBestFit}. The observational X-ray data are shown by the black points. Red dashed lines indicate the time of the neutron star's periastron.}
	\label{fig:XrayLuminosity}
\end{figure}

Figure \ref{fig:XrayLuminosity} shows the neutron star's mass capture converted into a luminosity and then attenuated by the additional local column
density of neutral hydrogen associated with the circumstellar disc (see Figure \ref{fig:NSObscuration}). The counts per second predicted by the simulations assumes instantaneous accretion of all matter captured by the neutron star. The simulation data replicates the times and magnitudes of the X-ray outbursts. It shows the X-ray count rate dropping to zero between the interactions of the neutron star with the Be star's circumstellar disc, whereas the observational X-ray data show non-zero count rates that steadily decrease over time. Thus, there is a reservoir of mass that fuels the X-ray flux of the neutron star around its orbit (i.e. a disc), which is fed by the Be star's disc during interaction. The interaction with the disc causes an increase of X-ray flux at binary phases of $\sim 0.1$ and $\sim 0.6$. As the disc shrinks, less matter is captured, leading to the decreasing trend in X-ray flux with time. The peaks in the simulation data seem to often precede those seen in the observational data which would also suggest a delay between mass capture and accretion, which has been suggested previously \citep{Okazaki2004,Brown2018}.

\section{H$\alpha$ profiles} \label{sec:Ha}

Be stars are well known for the H$\alpha$ emission produced by their discs \citep{RivCarc2013}. These profiles are the most direct method of observing the behaviour of the circumstellar disc. H$\alpha$ profiles provide information about the inclination of the system via the prominence of a double peaked structure that arises due to Doppler effects. The size of the disc can be inferred from the equivalent width of the H$\alpha$ profile. 

\citet{HorneMarsh1986} applied a simplification of radiative transfer techniques to model accretion discs, formulating the line emission shapes for optically thin and optically thick cases. This method has since been applied to Be star accretion discs by \citet{Okazaki1996} and \citet{HummelVrancken2000}. The method is unreliable for extreme inclination angles ($\geq$80$^{\circ}$). 

\citet{HorneMarsh1986} provided a simplified expression for the broadening of the line emission by Doppler shifts. These Doppler shifts are due to the orbital motion of the material in the disc along line of sight. This shear broadening is expressed as

\begin{equation} 
V_{\mathrm{shear}}(R) = -\frac{H}{2R} V_{K}(R) \sin i \tan i \sin \phi \cos \phi,
\label{eq:Vshear}
\end{equation}

\noindent where $i$ is the inclination of the disc, $\phi$ is the azimuthal angle in the disc plane, $H$ is the disc height, $R$ is the radius and $V_{K}$ is the local value of the Keplerian velocity. The line optical depth is given by

\begin{equation} 
\tau_{\nu} = \frac{W(R)}{\cos i} \frac{\lambda_{0}}{\sqrt{2 \pi} \Delta V} \exp \left[-\frac{1}{2} \left(\frac{V - V_{D}(0)}{\Delta V} \right)^{2} \right],
\label{eq:OD}
\end{equation}

\noindent where $\lambda_{0}$ is the rest wavelength, $V_{D}$ is the Doppler velocity shift produced by Keplerian motion and $\Delta V$ is given by

\begin{equation} 
\Delta V = \sqrt{\Delta V_{th}^{2} + V_{shear}^{2}}
\label{eq:deltaV}
\end{equation}

\noindent and $\Delta V_{th}$ is the thermal broadening. $W(R)$ is given by

\begin{equation} 
W(R) = \frac{\pi e^{2}}{m c}f\Sigma (R),
\label{eq:WR}
\end{equation}

\noindent where $f$ is the absorption oscillator strength and $\Sigma (R)$ is the surface density. In the previously mentioned treatments of shear broadening, the disc density is provided by a theoretical function. In this paper, the density is calculated at each integration step from the simulation particles.

\begin{figure}
	\centering
	\includegraphics[width=.5\textwidth]{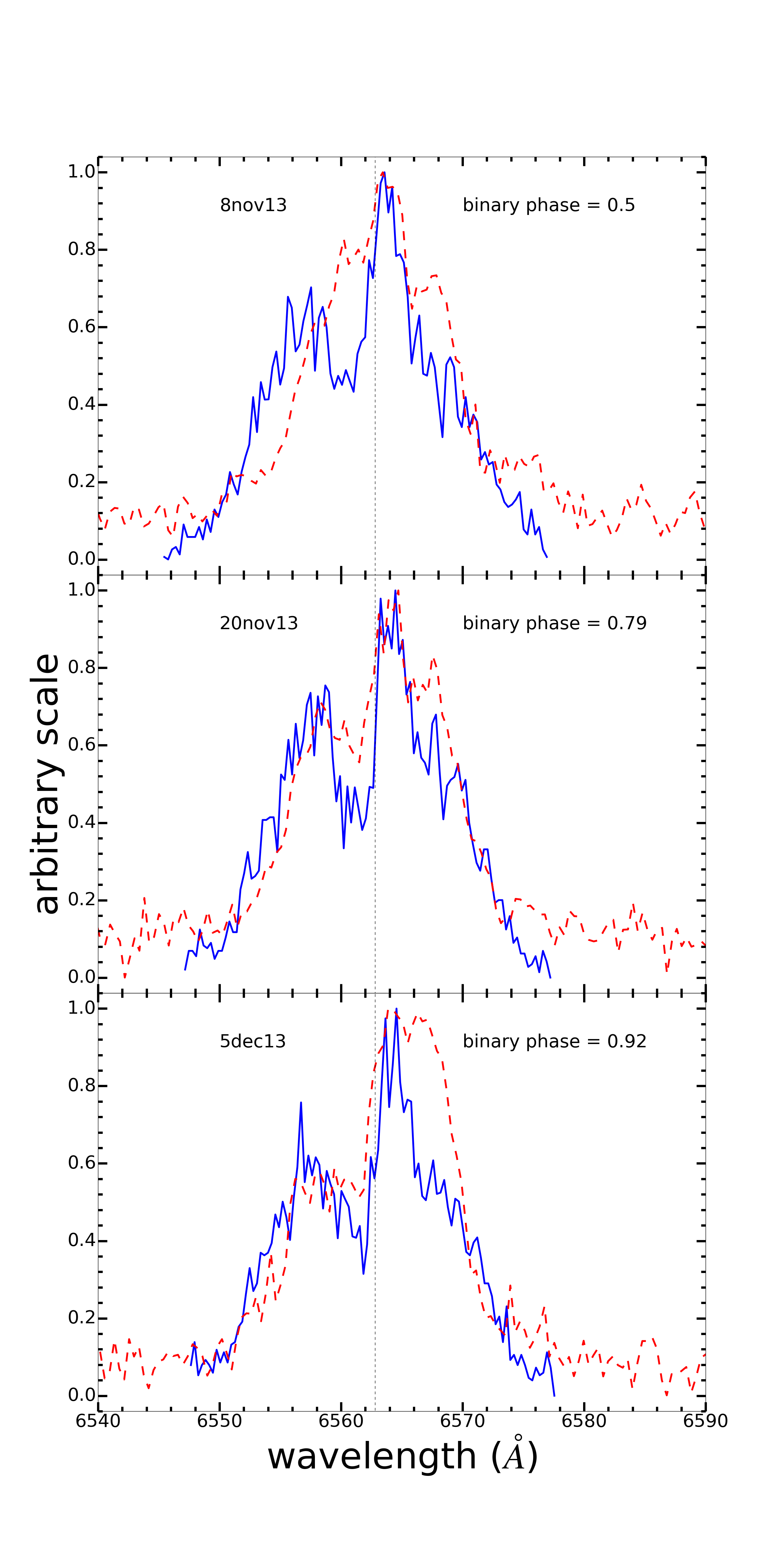} 
	\caption{Successive H$\alpha$ line emission shapes for the simulation data (in solid blue) compared against the observational line profiles (in dashed red) taken from \citet{Coe2015}. The profiles are produced at 5th November, 20th November and 5th December 2013. The black dashed line indicates the rest wavelength of H$\alpha$ emission, 6562.8\AA. The binary phase is shown for each profile, where a binary phase of 0.0 is periastron and 0.5 is apastron.}
	\label{fig:Halpha}
\end{figure}

Figure \ref{fig:Halpha} shows emission line profiles predicted by the simulation using the method described above. They are compared in the figure to the observational data from \citet{Coe2015} which have been corrected for the redshift of the SMC \citep{HarZar2006}. The key features of the observational data are replicated by the simulations. The H$\alpha$ profiles from the simulations exhibit a double peaked structure where the red (right) peak is larger, lies closer to the H$\alpha$ rest wavelength and has a shoulder on the redward side. This asymmetry can be attributed to the eccentricity of the disc, as discussed in Section \ref{sec:Eccentricity}. The profile has the worst fit at apastron (binary phase of 0.5) and the reverse is true for periastron. The model profiles show much smaller variations around an orbit than the observations.

\section{Discussion and Conclusions} \label{sec:Conclusions}

In this paper, a large optical outburst of the Be/X-ray binary, SXP 5.05, is investigated. Simulations are performed using the orbital solution derived in \citet{Coe2015}. As the most comprehensive set of observational data, the optical behaviour of the SXP 5.05 was used to narrow down the suitable simulations. The only way to fully replicate the observed behaviour of the optical data is to have a short and large increase in mass ejection initially, which then falls. The models struggle to reproduce the observational brightening after periastron and the amplitude of the variations. The use of radiative transfer methods is required to improve the accuracy of this work. The viscosity is also very high during the build up of the disc and lower during the dissipation - as described in \citet{Rimulo2018}. A mass ejection rate that rapidly decreases to the initial rate is assumed. The maximum value of the mass ejection is much larger than the expected range for mass ejection of Be stars and the assumed viscosity is larger than previously suggested but are similar to those used in \citet{Ghor2018} to model $\omega$ CMa.

The eccentricity of the disc begins very high at the beginning of the outburst and decreases over time. Different results would be acquired depending on when the outburst occurs during the Kozai-Lidov mechanism, due to the variation in the inclination and eccentricity of the Be star's circumstellar disc. As the eccentricity is cloase to $0.8$ at the beginning out the outburst, the Kozai-Lidov mechanicsm has taken effect. Investigating the possible role the K-L effect might play in such Type II outbursts is left for future work.

The increase in mass ejection causes the disc to increase in size, allowing it to grow larger than the truncation radius of the neutron star. Thus the neutron star is occulted at binary phases of $\sim 0.1$ and $\sim 0.6$: once as it passes behind the disc and once as it passes directly through it. The latter occultation is much smaller the the former. The measured column density of neutral hydrogen that obscures the neutron star allows the appropriate simulations to be constrained further. 

The X-ray flux of the simulations can be approximated by calculating the amount of mass captured by the neutron star and then applying a simple extinction function that is dependent on the aforementioned density of obscuring matter. The neutron star accretes at binary phases of $\sim 0.1$ and $\sim 0.6$. The time of the accretion and the attenuation of the X-ray flux matching the X-ray data from SXP 5.05. Hence, the X-ray behaviour of SXP 5.05 can be described well by the suggested model in this paper. The feature that the simulations do not model sufficiently, is the underlying X-ray flux seen between outbursts in the observational data. In the model, only the mass capture is used to calculate an instantaneous X-ray flux and so between accretion events, mass accretion drops to zero. The difference in behaviour could be explained by an accretion disc forming around the neutron star in SXP 5.05.

A simple treatment of the H$\alpha$ line profiles is applied to the modelled disc and yields all the general features of the profiles seen in \citet{Coe2015}. The red peak lies close to the rest wavelength and is much larger than the blue. The asymmetry of the predicted H$\alpha$ profiles is due to the large eccentricity of the Be star's circumstellar disc. There is even a shoulder to the right of the red peak. However, this method does not recover the variations with time seen in the observational data. 

Optical variations in Be star systems are be due to changes in the structure and size of the disc. The evolution of the disc is itself dependent on the mass ejection which is, in turn, dependent on the non-radial pulsations of the star. These non-radial pulsations are inherent to the star and do not change. They typically possess periods of the order of days. Thus, the large optical event seen in SXP 5.05 could be the result of the overlap of a number of the non-radial pulsations and hence, this event has a periodicity to it. 

The results in this paper imply a very specific nature to optical outbursts that occur in Be stars; a very sharp and large increase in the amount of matter ejected followed by an rapid decline over time. The data also implies that any system with a disc heavily inclined to the orbital plane and undergoing a sufficiently large outburst will possess a disc that grows outside of the orbit of the compact object. This can be tested by the presence of an additional occultation of the X-ray source during the outburst. In the case of SXP 5.05, which possesses one obscuration of the neutron star per orbit and an additional one during outburst, it is possible to constrain the inclination angle to the observer. Therefore, this should be possible for similar Be/X-ray binaries.


\section*{Acknowledgements}

We thank Vanessa McBride for providing us with copies of the H$\alpha$ spectra shown in Figure \ref{fig:Halpha}. ROB thanks Alex Carciofi and Amanda Rubio for their helpful discussions. ROB acknowledges support from the Engineering and Physical Sciences Research Council Centre for Doctoral Training grant EP/L015382/1. WCGH acknowledges support from the Science and Technology Facilities Council through grant number ST/M000931/1. The authors acknowledge the use of the IRIDIS High Performance Computing Facility, and associated support services at the University of Southampton, in the completion of this work. Numerical computations were also performed on the Sciama High Performance Compute (HPC) cluster which is supported by the ICG, SEPNet and the University of Portsmouth.





\begin{thebibliography}{99}

\bibitem[\protect\citeauthoryear{Brown et al.}{2018}]{Brown2018}
Brown R. O., Ho W. C. G., Coe M. J. and Okazaki A. T., 2018, MNRAS, 477, 4810

\bibitem[\protect\citeauthoryear{Carciofi}{2011}]{Carciofi2011} 
Carciofi A. C., 2011, Active OB Stars: Structure, Evolution, Mass Loss, and Critical Limits, 325, IAUS..272
\bibitem[\protect\citeauthoryear{Carciofi and Bjorkman}{2006}]{CarcBjor2006}
Carciofi A. C. and Bjorkman J. E., 2006, \apj, 639, 1081
\bibitem[\protect\citeauthoryear{Coe et al.}{2013}]{Coe2013}
Coe M. J. et al., 2013, The Astronomer's Telegram, No. 5547
\bibitem[\protect\citeauthoryear{Coe et al}{2015}]{Coe2015}
Coe M. J. et al., 2015, MNRAS, 447, 2387
\bibitem[\protect\citeauthoryear{Coleiro et al.}{2013}]{Coleiro2013}
Coleiro A. et al., 2013, A\&A, 560, A108

\bibitem[\protect\citeauthoryear{Fu, Lubow and Martin}{2015}]{Fu2015}
Fu W., Lubow S. H. and Martin R. B., 2015, Astrophys. J., 807, 14
\bibitem[\protect\citeauthoryear{Fu, Lubow and Martin}{2017}]{Fu2017}
Fu W., Lubow S. H. and Martin R. B., 2015, Astrophys. J., 835, 5

\bibitem[\protect\citeauthoryear{Ghoreyshi et al.}{2018}]{Ghor2018}
Ghoreyshi M. R. et al., 2018, MNRAS, 479, 2214

\bibitem[\protect\citeauthoryear{Harris and Zaritsky}{2006}]{HarZar2006}
Harris J. and Zaritsky D., 2006, \aj, 131, 2514 
\bibitem[\protect\citeauthoryear{Haubois et al.}{2012}]{Haubois2012}
Haubois X., Carciofi A. C., Rivinius T., Okazaki A. T., Bjorkman J. E., 2012, Astrophys. J., 756, 156
\bibitem[\protect\citeauthoryear{Horne and Marsh}{1986}]{HorneMarsh1986}
Horne, K. and Marsh T. R., 1986, Monthly Notices of the Royal Astronomical Society, 218, 761
\bibitem[\protect\citeauthoryear{Hummel and Vrancken}{2000}]{HummelVrancken2000}
Hummel, W. and Vrancken, M., 2000, A\&A, 359, 1075

\bibitem[\protect\citeauthoryear{Jaschek and Egret}{1982}]{JascEgre1982}
Jaschek M. and Egret D., 1982, Proceedings IAU Symposium No. 98, Dordrecht: D. Reidel Publishing Co., 261

\bibitem[\protect\citeauthoryear{King et al.}{2013}]{King2013}
King A. R., Livio M., Lubow S. H. and Pringle J. E., 2013, MNRAS, 431, 2655

\bibitem[\protect\citeauthoryear{Martin et al.}{2014}]{Martin2014} 
Martin R. G., Nixon C., Lubow S. H., Armitage P. J., Price D. J., Do{\u g}an S., King A., 2014, \apj, 792, 33

\bibitem[\protect\citeauthoryear{Okazaki}{1996}]{Okazaki1996}
Okazaki A. T., 1996, Publications of the Astronomical Society of Japan, 48, 305
\bibitem[\protect\citeauthoryear{Okazaki and Hayasaki}{2004}]{Okazaki2004}
Okazaki A. T. and Hayasaki K., 2004, Revista Mexicana de Astronomia y Astrofisica (Serie de Conferencias), 20, 144
\bibitem[\protect\citeauthoryear{Okazaki et al}{2002}]{Okazaki2002}
Okazaki A. T., Bate M. R., Ogilvie G. I. and Pringle J. E., 2002, MNRAS, 337, 967

\bibitem[\protect\citeauthoryear{Poeckert and Marlborough}{1982}]{PoeMarl1982}
Poeckert R. and Marlborough J. M., 1982,  Astrophys. J., 252, 196

\bibitem[\protect\citeauthoryear{Rappaport and van de Heuvel}{1982}]{RapHeu1982}
Rappaport S. and van de Heuvel E. P. J., 1982, IAU Symposium  No. 98 "Be Stars" (M. Jaschek and H.G. Groth, Editors), Reidel, Dordrecht, 327
\bibitem[\protect\citeauthoryear{Reig et al.}{2011}]{Reig2011}
Reig, P.\ 2011, \apss, 332, 1
\bibitem[\protect\citeauthoryear{R\'{i}mulo et al.}{2018}]{Rimulo2018}
R\'{i}mulo L. R. et al., 2018, MNRAS, 476, 3555
\bibitem[\protect\citeauthoryear{Rivinius}{2000}]{Riv2000}
Rivinius T., 2000, ASP Conference Series, 214, 228
\bibitem[\protect\citeauthoryear{Rivinius, Carciofi and Martayan}{2013}]{RivCarc2013}
Rivinius T., Carciofi A. C. and Martayan C., 2013, A\&A Review, 21, 69

\bibitem[\protect\citeauthoryear{Scowcroft}{2016}]{Scowcroft2016}
Scowcroft V. et al., 2016, \apj, 816, 49 
\bibitem[\protect\citeauthoryear{Slettebak}{1988}]{Slettebak1988}
Slettebak A., 1988, Publ. Astron. Soc. Pac, 100, 770

\bibitem[\protect\citeauthoryear{van de Heuvel and Rappaport}{1987}]{HeuRap1987}
van den Heuvel E.P.J. and Rappaport S., 1987, IAU Colloquium No. 92 "Physics of Be Stars" (A. Slettebak and T.P. Snow, editors), Cambridge University Press, 291
\bibitem[\protect\citeauthoryear{van Kerkwijk et al.}{1995}]{vanKer1995}
van Kerkwijk M. H., Waters L. B. F. M. and Marlborough J. M., 1995, A\&A, 300, 259
\bibitem[\protect\citeauthoryear{Vieira et al.}{2015}]{Vieira2015}
Vieira R. G. et al., 2015, MNRAS, 454, 2107

\end{thebibliography}








\label{lastpage}
\end{document}